\documentclass[12pt,a4paper]{article}
\usepackage[top=2.1cm,bottom=2.1cm,left=2.1cm,right=2.1cm]{geometry}
\usepackage{latexsym,amsmath,amsfonts,amssymb,amsthm,amscd,mathrsfs}

\usepackage{accents}

\newcommand{\be}{\begin{equation}}
\newcommand{\ee}{\end{equation}}
\newcommand{\bea}{\begin{eqnarray}}
\newcommand{\eea}{\end{eqnarray}}

\numberwithin{equation}{section}

\numberwithin{thmcounter}{section}

\theoremstyle{definition}
\newtheorem*{acknowledgements}{Acknowledgements}

\theoremstyle{plain}

\def\1{{\boldsymbol 1}}                     %
\def\cD{{\mathcal D}}                       %
\def\cH{{\mathcal H}}                       %
\def\cP{{\mathcal P}}                       %
\def\tr{\mathrm{tr}}                        %
\def\diag{\mathrm{diag}}                    %
\def\ri{{\rm i}}                            %
\def\C{\mathbb{C}}                          %
\def\N{\mathbb{N}}                          %
\def\R{\mathbb{R}}                          %
\def\T{\mathbb{T}}                          %
\def\UN{{\rm U}}                            %
\def\GL{{\rm GL}}                           %
\def\fH{\mathfrak{H}}                       %
\def\cF{{\mathcal F}}                       %
\def\fP{{\mathfrak{P}}}                     %
\def\reg{\mathrm{reg}}                      %
\def\red{\mathrm{red}}                      %
\def\span{{\mathrm{span}}}                  %
\def\cR{{\mathcal R}}                       %
\def\Ad{{\mathrm{Ad}}}                      %
\def\id{{\mathrm{id}}}                      %
\def\cA{{\mathcal A}}                       %
\def\dt {\left.\frac{d}{dt}\right|_{t=0}}   %
\def\Herm{{\mathfrak{H}}}                   %
\def\fM{\mathfrak{M}}                       %
\def\u{\mathfrak{u}}                        %
\def\b{\mathfrak{b}}                        %
\def\B{\mathrm{B}}                          %
\def\gl{\mathfrak{gl}(n,\C)}                %
\def\cN{{\mathcal N}}                       %

\begin{document}

\begin{center}
{\Large\bf
Reduction of a  bi-Hamiltonian hierarchy on
$T^*\UN(n)$ to spin Ruijsenaars--Sutherland  models}
\end{center}

\medskip
\begin{center}
L.~Feh\'er${}^{a,b}$
\\

\bigskip
${}^a$Department of Theoretical Physics, University of Szeged\\
Tisza Lajos krt 84-86, H-6720 Szeged, Hungary\\
e-mail: lfeher@physx.u-szeged.hu

\medskip
${}^b$Department of Theoretical Physics, WIGNER RCP, RMKI\\
H-1525 Budapest, P.O.B.~49, Hungary\\
\end{center}

\medskip
\begin{abstract}
We first exhibit two compatible Poisson structures on the cotangent bundle of the unitary group
 $\UN(n)$
in such a way that the invariant functions of the $\mathfrak{u}(n)^*$-valued
momenta generate a bi-Hamiltonian hierarchy. One of the Poisson structures is the
canonical one and the other one arises from embedding the Heisenberg double of the
Poisson-Lie  group $\UN(n)$ into $T^*\UN(n)$, and subsequently extending
the embedded Poisson structure to the full cotangent bundle.
We then apply Poisson reduction
to the bi-Hamiltonian hierarchy on $T^* \UN(n)$ using the conjugation action of $\UN(n)$, for which
the ring of invariant functions is closed under both Poisson brackets.
We demonstrate that the reduced hierarchy belongs to the overlap of well-known
trigonometric  spin Sutherland and
spin Ruijsenaars--Schneider type integrable many-body models,
which receive a bi-Hamiltonian interpretation via our treatment.
\end{abstract}

{\linespread{0.8}\tableofcontents}

\newpage
\section{Introduction}
\label{sec:I}

The many-body models of Calogero--Moser--Sutherland  \cite{Cal,M,S} and Ruijsenaars--Schneider \cite{RS} type are
among the most interesting examples of finite-dimensional integrable systems
both from the mathematical point of view and regarding their diverse physical applications.
See, for example,  the  reviews \cite{N,RBanff,vDV} and references therein.
The spin extensions of these models \cite{GH,KZ,W} are also important,
and are currently subject to intense studies \cite{AO,CF2,F1,F2,KLOZ,Res2,Res3,SS}.

The purpose of the present paper is to deepen the understanding of the Hamiltonian
structure for  a system
of evolution equations that belongs to the above-mentioned family.
The equations at issue have the form
\be
\dot{Q}= (\ri L^k)_0 Q,
\qquad
\dot{L} = [ \cR(Q)( \ri L^k), L],
\label{I1}\ee
where $Q\in \T^n_\reg$ is a diagonal unitary matrix with distinct eigenvalues,
$L\in \Herm(n)$ is an $n\times n$ Hermitian matrix,
and the subscript $0$ means diagonal part.
 The dynamical $r$-matrix $\cR(Q)$
is the linear operator on $\gl$ that acts as zero on the diagonal matrices and acts on
off-diagonal matrices according to
\be
\cR(Q) = \frac{1}{2} (\Ad_Q + \id)(\Ad_Q - \id)^{-1}.
\label{I2}\ee
The inverse is well-defined on the off-diagonal subspace
 by virtue of the regularity of $Q$;
$\Ad_Q(X) = Q X Q^{-1}$.
The evolutional derivations (\ref{I1}) associated with arbitrary $k\in \N$ mutually
commute \emph{if} one restricts
attention to `observables' $f(Q,L)$ that are invariant with respect to conjugations of $L$
by diagonal unitary matrices.
This may motivate one to identify the phase space of the system as one of the two quotient
spaces:
\be
\T^n_\reg \times \left(\Herm(n)/\T^n\right)
\quad
\hbox{or}\quad
\left(\T^n_\reg \times \Herm(n)\right)/\cN(n),
\label{I3}\ee
where $\cN(n)$ is the normalizer of the maximal torus $\T^n< \UN(n)$ in $\UN(n)$,
acting by simultaneous conjugations of both $Q$ and $L$.
The latter choice is actually more natural since it takes into account a hidden symmetry
with respect to the permutation group $S_n = \cN(n)/\T^n$.
Accordingly, the physical observables are identified as the invariant real functions forming
\be
C^\infty(\T^n_\reg \times \Herm(n))^{\cN(n)}.
\label{I4}\ee
The system has a well-known Hamiltonian structure \cite{EFK,FP1,LX,Res1}, which arises
via the parametrization
\be
L= p - (\cR(Q) + \frac{1}{2}\id)(\phi),
\label{I5}\ee
where $p\in \Herm(n)_0$ and $\phi\in \Herm(n)_\perp$, that is,
they are Hermitian diagonal and off-diagonal matrices, respectively.
The diagonal entries $p_j$ of $p$ and $q_j$ in $Q_j = e^{\ri q_j}$ represent canonically
conjugate pairs, and
are combined with the Poisson algebra carried by the quotient
\be
\Herm(n)_\perp /\T^n \equiv \u(n)^*/\!/_0 \T^n.
\label{I6}\ee
The quotient (\ref{I6}) embodies a Hamiltonian reduction
 \cite{OR} of the Lie-Poisson bracket of $\u(n)$ defined by
 utilizing the  action of $\T^n< \UN(n)$  on  $\u(n)^*\equiv \Herm(n)$.
In correspondence with (\ref{I4}), only the $S_n$-invariant elements of the full
Poisson algebra are kept.
The $k=1$ member of the `hierarchy' (\ref{I1}) is generated by the
 standard spin Sutherland Hamiltonian
\be
H_{\mathrm{Suth}}(Q,p,\phi)= \frac{1}{2} \tr \left(L(Q,p,\phi)^2\right)   =
\frac{1}{2} \sum_{i=1}^n p_i^2 + \frac{1}{8} \sum_{k\neq l}
\frac{\vert \phi_{kl}\vert^2}{\sin^2\frac{q_k - q_l}{2}}.
\label{I7}\ee
We stress that the Hamiltonian belongs to the space (\ref{I4}) and  governs the time development
of the physical observables.

In the `unparametrized form' (\ref{I1}) the system can be viewed also as a degenerate limiting
case of the spin Ruijsenaars--Schneider (RS) models introduced by Krichever and Zabrodin \cite{KZ}.
This interpretation was pointed out in the papers \cite{BH,Li}, without noticing
the coincidence  with the spin Sutherland model.
To be more exact,
in these references the hyperbolic analogue of  the system (\ref{I1}) was
considered.

It is well-known that the restriction  of the system (\ref{I1}) to a $2n$-dimensional
symplectic leaf
of the above-mentioned Poisson structure gives
the spinless Sutherland model \cite{KKS}.
Interestingly, as explained below,   another
specialization gives the spinless trigonometric RS model.
This latter specialization arises by restriction to a $2n$-dimensional symplectic leaf
with respect to another Poisson structure, from which the same equations can be derived.
To emphasize its double interpretation, the system (\ref{I1}) will be referred to as
the trigonometric \emph{spin Ruijsenaars--Sutherland hierarchy}.

The standard Poisson structure of the spin Sutherland model (\ref{I7}) results
  by applying a Poisson reduction \cite{FP1,KKS,Res1} to the canonical Poisson structure
of the cotangent bundle $T^*\UN(n)$.   Our principal goal is to show that the
cotangent bundle can be equipped with another Poisson structure, too, whose
reduction induces another Poisson bracket on the space of
observables (\ref{I4}). \emph{The two Poisson
structures on $T^*\UN(n)$ as well as their reductions to (\ref{I4}) turn out to
be compatible in the
sense of bi-Hamiltonian geometry, and the evolution equations
(\ref{I1}) as well as their unreduced avatars enjoy
the bi-Hamiltonian property.} This is the main result of the paper.
(For background on bi-Hamiltonian systems, see~e.g.~\cite{FMP,Se,Sm}.)
As we shall see,
the pertinent second Poisson structure is transferred to the cotangent bundle
 from the Heisenberg double \cite{STS} of the  Poisson-Lie group $\UN(n)$.

The spinless trigonometric RS model was derived in \cite{FK1,FK2}
by symplectic reduction of the free system on the Heisenberg double of $\UN(n)$ at a
particular value of the corresponding moment map.
It is true in general that the reduced phase spaces of symplectic reduction
are  symplectic leaves in the quotient of the original phase space
defined by Poisson reduction. This explains how the spinless RS model appears
on a symplectic leaf of the Ruijsenaars--Sutherland hierarchy with respect
to its second Poisson structure.
The reader may consult \cite{F1}, too,
where, we studied symplectic reductions of Heisenberg doubles at arbitrary moment map values;
but without dealing with any bi-Hamiltonian aspect.

The bi-Hamiltonian structure of the hyperbolic analytic continuation of the
trigonometric system (\ref{I1}) is described in \cite{F2}. However, in that case we do not
have an explanation via a single Poisson reduction.  Incidentally, a related problem is
that
no Hamiltonian reduction treatment
of the real, repulsive
hyperbolic spinless RS model is known\footnote{This is so despite the fact that the
holomorphic hyperbolic/trigonometric  RS model is well-understood in more than one
reduction approaches \cite{AKO,CF1,FR,Ob}.
Treating real forms of holomorphic
integrable systems is a highly non-trivial task in general.}.

Now we give an outline of the rest of the text.
We start in Section 2  by presenting a tailor-made account of the Heisenberg double.
In Section 3,  we exhibit the bi-Hamiltonian structure on the
cotangent bundle, and show that the free geodesic motion on $\UN(n)$
is encoded by
a bi-Hamiltonian system.
This is the content of Proposition 3.2 together with Lemma 3.3, which
represent our first new result.
Section 4 is the essential part of the paper, where we characterize the
Poisson reduction of the bi-Hamiltonian manifold $\fM:=T^*\UN(n)$.
Our main result is Theorem 4.5, which gives the compatible
Poisson brackets on the space of functions (\ref{I4}).
In addition, we show that the equations of motion (\ref{I1}) descend
from the free system on $T^* \UN(n)$, and also display a large
set of constants of motion.
In Section 5, we give our conclusions and
 discuss how spin degrees of freedom can be introduced
in relation to the second Poisson bracket.

\section{The rudiments of the Heisenberg double}
\label{sec:K}

The material collected below is well known to experts (see e.g. \cite{FK2, Lu,Kli,KS,STS}),
except perhaps the presentation of the quasi-adjoint action that we shall give.
We start by recalling that the Heisenberg double of the standard Poisson-Lie group $\UN(n)$
is the real Lie group $\GL(n,\C)$ equipped with a certain Poisson structure.
This Poisson structure is actually symplectic, and it contains all information
about the Poisson structure on $\UN(n)$ as well.

Before presenting the Poisson structure, we introduce  two diffeomorphisms
\be
m_1: \GL(n,\C) \to \UN(n) \times \B(n),
\quad
m_2: \UN(n) \times \B(n) \to \UN(n) \times \fP(n),
\label{K1}\ee
where $\B(n)$ is the subgroup of $\GL(n,\C)$ consisting of the upper triangular
matrices with positive diagonal entries,
and $\fP(n)$  contains the Hermitian, positive elements of $\GL(n,\C)$.
Every element $K\in \GL(n,\C)$
admits the unique decompositions
\be
K = b_L g_R^{-1} = g_L b_R^{-1}\quad\hbox{with}\quad b_L, b_R \in \B(n),\, g_L, g_R\in \UN(n),
\label{Iwas}\ee
and $K$ can be recovered also from the pairs $(g_L, b_L)$ and $(g_R, b_R)$,
by utilizing the decompositions
\be
b_L^{-1} g_L = g_R^{-1} b_R.
\label{K3}\ee
It is easily seen from this that the map $m_1$ defined by
\be
m_1(K) := (g_R, b_R)
\label{K4}\ee
is a diffeomorphism; and so is the map
\be
m_2(g_R, b_R):= (g_R, b_R b_R^\dagger).
\label{K5}\ee
We shall use these maps to transfer the Poisson structure of
$\GL(n,\C)$ to the model spaces $\UN(n) \times \B(n)$ and $\UN(n) \times \fP(n)$.

Consider the \emph{real} Lie algebra $\gl$ and equip it
with the non-degenerate, invariant bilinear form
\be
\left\langle X,Y \right\rangle:= \Im\tr(XY),
\quad\forall X,Y\in \gl.
\label{form}\ee
Introduce the linear subspace of Hermitian matrices
\be
\Herm(n):= \ri \u(n),
\label{K7}\ee
and the subalgebra
\be
\b(n):= \span_\R\{ E_{jj}, E_{kl}, \ri E_{kl} \mid 1\leq j\leq n,\,\, 1\leq k < l\leq n\},
\label{K8}\ee
where  $E_{kl}$ is the elementary matrix of size $n$, having $1$ at the $kl$ position.
Both $\Herm(n)$ and $\b(n)$ are in duality with $\u(n)$ with respect to the
bilinear form (\ref{form}). The real vector space decomposition
\be
\gl = \u(n) + \b(n)
\label{dec+}\ee
allows us to write every $X\in \gl$ in the form
\be
X= X_{\u(n)} + X_{\b(n)}
\label{Xdec}\ee
with constituents in the respective subalgebras.

For any real function\footnote{If not specified otherwise,
our spaces of $C^\infty$-functions always denote spaces of \emph{real} functions.}
$f \in C^\infty(\GL(n,\C))$,  define the $\gl$-valued
derivatives $\nabla f$ and $\nabla' f$ by
\be
\langle \nabla f(K), X\rangle := \dt f(e^{tX} K),
\quad
\langle \nabla' f(K), X\rangle := \dt f(Ke^{tX}),
\,\, \forall X\in \gl.
\label{K11}\ee
For any function $\phi \in C^\infty(\UN(n))$  introduce the
$\b(n)$-valued derivatives by
\be
\langle D \phi(g), X\rangle := \dt \phi(e^{tX} g),
\quad
\langle D' \phi(g), X\rangle := \dt \phi(ge^{tX}),
\,\, \forall X\in \u(n),
\label{K12}\ee
and for any $\chi \in C^\infty(\B(n))$ introduce the $\u(n)$-valued derivatives by
\be
\langle D \chi(b), X\rangle := \dt \chi(e^{tX} b),
\quad
\langle D' \chi(b), X\rangle := \dt \chi(be^{tX}),
\,\, \forall X\in \b(n).
\label{K13}\ee
Finally, for $\psi \in C^\infty(\fP(n))$, define the $\u(n)$-valued
derivative $d \psi$ by
\be
\langle d \psi(L), X\rangle := \dt \psi(L + t X),
\,\,\forall X\in \Herm(n).
\label{K14}\ee
This definition makes sense since $(L + t X)\in \fP(n)$ for small $t$; remember that
$\fH(n):= \ri \u(n)$.

Following Semenov-Tian-Shansky \cite{STS}, we introduce
the  (non-degenerate) Poisson bracket  $\{\ ,\ \}_+$ on  $C^\infty(\GL(n,\C))$ by
\be
\{ f_1, f_2\}_+ = \langle \nabla f_1, R \nabla f_2 \rangle +  \langle \nabla' f_1, R \nabla' f_2 \rangle,
\label{+PB}\ee
where $R := \frac{1}{2}\left( P_{\u(n)} - P_{\b(n)}\right)$ is half the difference of the projection
operators on $\gl$ associated with the decomposition (\ref{dec+}).

We can express the Poisson structure of the Heisenberg double in terms of the
variables $(g, b)\equiv (g_R, b_R)$ and $(g,L)\equiv (g_R, b_R b_R^\dagger)$.
In other words, the manifolds
$\UN(n) \times \B(n)$ and $\UN(n) \times \fP(n)$ carry unique Poisson structures $\{\ ,\ \}_+^1$
and  $\{\ ,\ \}_+^2$ for which
\be
m_1: \left(\GL(n,\C), \{\ ,\ \}_+\right) \to \left(\UN(n) \times \B(n),\{\ ,\ \}_+^1\right)
\label{m1}\ee
and
\be
m_2: \left(\UN(n) \times \B(n), \{\ ,\ \}_+^1\right) \to
\left(\UN(n)\times \fP(n),\{\ ,\ \}^2_+\right)
\label{m2}\ee
are Poisson diffeomorphisms.
Straightforward calculations lead to the following formulas.

\medskip
\noindent
{\bf Proposition 2.1.}
\emph{For
$\cF\in C^\infty(\UN(n) \times \B(n))$ denote $D_1\cF$ and $D_2\cF$ the
derivatives with respect to the first and second arguments.
The Poisson bracket of $\cF, \cH \in C^\infty(\UN(n)\times \B(n))$ can be written as follows:
\bea
&&\{\cF, \cH\}_+^1(g,b) =\left\langle D_2' \cF, b^{-1} (D_2\cH) b \right\rangle
-\left\langle D'_1\cF, g^{-1} (D_1\cH) g\right\rangle
\nonumber\\
&&\qquad \qquad  +  \left\langle D_1\cF , D_2\cH \right\rangle
-\left\langle D_1 \cH , D_2\cF \right\rangle,
\label{+PB1}\eea
where the derivatives on the right-hand side
are   taken at $(g,b)\in \UN(n)\times \B(n)$.}

\begin{proof}
Take an arbitrary function $\cH \in C^\infty(\UN(n) \times \B(n))$ and the corresponding function $h= m_1^*(\cH)$.
The claim is equivalent to the statement that the Hamiltonian vector field of $\cH$ is the push-forward
of the Hamiltonian vector field of $h$ by the diffeomorphism $m_1$.
Denote the Hamiltonian vector field of $h$ simply by dot.
From the formula (\ref{+PB}), we get
\be
\dot{K} = (R \nabla h(K)) K + K (R\nabla' h(K)).
\label{dotK}\ee
Consider the special case when
\be
h(K) = \psi(b_R) \quad\hbox{for some}\quad \psi\in C^\infty(\B(n)),
\ee
i.e., $h = m_1^*(\cH)$ for the function $\cH$ given by $\cH(g,b) = \psi(b)$.
A routine calculation shows that the derivatives of $h$ are related to $D'\psi$ by
\be
\nabla' h(K) = -b_R \left( D' \psi(b_R) \right) b_R^{-1},
\qquad
\nabla h(K) =  - g_L \left(D' \psi(b_R)\right) g_L^{-1}.
\ee
Plugging this into (\ref{dotK}) and using that $D\psi(b_R) = \left(b_R (D' \psi(b_R)) b_R^{-1} \right)_{\u(n)}$, we obtain
\be
\dot{K} = -K D\psi(b_R).
\label{dotK2}\ee
The decomposition $K= b_L g_R^{-1}$ and  the formula (\ref{dotK2})  give
\be
\dot{b}_L =0,\qquad  \dot{g}_R = (D\psi(b_R)) g_R.
\label{dotbLgR}\ee
On the other hand, the decomposition $K= g_L b_R^{-1}$ and (\ref{dotK2}) give
\be
\dot{K} = g_L ( g_L^{-1} \dot{g}_L - b_R^{-1} \dot{b}_R) b_R^{-1} = - g_L \left(b_R^{-1} (D\psi(b_R)) b_R \right) b_R^{-1},
\ee
and this implies
\be
\dot{b}_R = b_R \left( b_R^{-1} (D\psi(b_R)) b_R\right)_{\b(n)}, \qquad
\dot{g}_L  = - g_L D'\psi(b_R).
\label{dotbRgL}\ee
By combining (\ref{dotbLgR}) and (\ref{dotbRgL}),
we conclude
that the push-forward of the Hamiltonian vector field
of the function $h$ is encoded by the formula
\be
\dot{b} = b (b^{-1} \left(D\psi(b)) b\right)_{\b(n)}, \qquad
\dot{g} = (D\psi(b))g.
\label{V2}\ee
Clearly, this agrees with the Hamiltonian vector field associated with the function $\cH $
by means of the bracket (\ref{+PB1}).

A similar calculation proves the desired result
for functions of the form $h(K) = \phi(g_R)$.
By the properties of bi-derivations\footnote{Our calculations do not use the
Jacobi identity of the $m_1$-related brackets. Hence the Jacobi identity of
the bracket $\{\ ,\ \}_+^1$ (\ref{+PB1})
follows from that of $\{\ ,\ \}_+$ (\ref{+PB}).},
the required statement about the Hamiltonian vector fields of $\cH$ and $m_1^*(\cH)$
then holds for arbitrary functions
of the form $\cH(g,b)  = \psi(b) \phi(g)$ as well.
It is easy to see that this is sufficient for establishing the validity of  the
proposition.
\end{proof}

\medskip\noindent
{\bf Proposition 2.2.}
\emph{ For
$F\in C^\infty(\UN(n) \times \fP(n))$ denote $D_1 F$ and $d_2 F$ the
derivatives with respect the first and second arguments.
We have the following formula:
\bea
&&\{F, H\}_+^2 (g,L)=4\left\langle L d_2F, \left(L d_2 H \right)_{\u(n)}\right\rangle
-\left\langle D'_1 F , g^{-1} (D_1 H) g  \right\rangle\nonumber \\
&&\qquad \qquad  +  2\left\langle D_1F, L d_2H \right\rangle
-2\left\langle D_1 H, L d_2F\right\rangle,
\label{+PB2}\eea
where the derivatives are taken at $(g,L)\in \UN(n)\times \fP(n)$,
and (\ref{Xdec}) is applied to $X= (L d_2 H)$.}

\begin{proof}
Notice that any $\cF\in C^\infty(\UN(n)\times \B(n))$ corresponds to a unique
$F\in C^\infty(\UN(n) \times \fP(n))$ according to
\be
\cF(g,b) = F(g,L)
\quad\hbox{with}\quad L=bb^\dagger.
\ee
For arbitrarily chosen $(g,b)$, $X\in \b(n)$ and $Y\in \u(n)$,  consider the derivatives of $\cF$
and $F$ at $t=0$ along the curves $(g, b(t))$
and $(g, L(t))$ with $L(t) = b(t) b(t)^\dagger$ using
\be
b(t) = \exp(tX) b
\quad\hbox{and}\quad
b(t) = b\exp\left( t(b^{-1} Y b)_{\b(n)}\right).
\ee
By elementary manipulations, the equality of the derivatives gives
\be
\langle X+Y, b (D_2' \cF(g,b)) b^{-1} -  2 L d_2 F(g,L)\rangle =0,
\ee
which implies
\be
b \left(D_2' \cF(g,b) \right) b^{-1} =  2 L d_2 F(g,L).
\ee
Relying on this and the obvious relations
\be
D_1 \cF(g,b) = D_1 F(g,L),
\quad
D_2 \cF(g,b) = \left(b (D_2' \cF(g,b)) b^{-1}\right)_{\u(n)},
\ee
we can convert formula (\ref{+PB1}) into (\ref{+PB2}).
\end{proof}

Referring to the decompositions (\ref{Iwas}), let us now introduce the maps $\Lambda_L, \Lambda_R$
 from $\GL(n,\C)$ to $\B(n)$,
and the maps $\Xi_L, \Xi_R$ from $\GL(n,\C)$ to $\UN(n)$ by
\be
\Lambda_L(K):= b_L,\, \Lambda_R(K):= b_R,\quad \Xi_L(K):= g_L,\, \Xi_R(K):= g_R.
\label{K20}\ee
The maps $\Lambda_L$ and  $\Lambda_R$ are Poisson map with respect to the standard
multiplicative Poisson bracket on $\B(n)$,
which is encoded by the first term of the formula (\ref{+PB1}). Moreover, the map
$\Lambda:= \Lambda_L \Lambda_R: \GL(n,\C) \to \B(n)$, given by
\be
\Lambda(K):= b_L b_R,
\label{K21}\ee
is also a Poisson map. Similarly, the maps $\Xi_L$ and $\Xi_R$ are Poisson maps,
if $\UN(n)$ is endowed with the Poisson structure that appears in the second term of (\ref{+PB1}).
It follows from general results that $\Lambda$ is the moment map, in the sense of Lu
\cite{Lu}, for a certain
Poisson action of $\UN(n)$ on the Heisenberg double.
This action was named `quasi-adjoint action' by Klim\v c\'\i k \cite{Kli}.

For any $\eta\in \UN(n)$,
let $\cA_\eta$ denote the diffeomorphism of $\GL(n,\C)$ associated with the quasi-adjoint action.
It operates \cite{Kli} according to
\be
\cA_\eta(K) = \eta K \Xi_R( \eta \Lambda_L(K) ).
\label{K22}\ee
The quasi-adjoint Poisson action
\be
\cA: \UN(n) \times \GL(n,\C) \to \GL(n,\C),
\quad
\cA(\eta,K):= \cA_\eta(K),
\label{K23}\ee
gives rise to Poisson actions $\cA^1$ and $\cA^2$ on $\UN(n) \times \B(n)$ and on
$\UN(n) \times \fP(n)$, respectively,   via the definitions
\be
\cA^1_\eta := m_1 \circ \cA_\eta \circ m_1^{-1}
\quad\hbox{and}\quad
\cA^2_\eta:= m_2 \circ \cA_\eta^1 \circ m_2^{-1}.
\label{K24}\ee
One can check that these actions obey the following formulas:
\be
\cA_\eta^1(g,b) =  (\tilde \eta g \tilde \eta^{-1}, \Lambda_L( \tilde \eta b))
\quad\hbox{with}\quad \tilde \eta = \Xi_R(\eta \Lambda_L(m_1^{-1}(g,b)))^{-1}
\label{K25}\ee
and
\be
\cA_\eta^2(g,L) =  (\tilde \eta g \tilde \eta^{-1}, \tilde \eta L \tilde \eta^{-1})
\quad\hbox{with}\quad \tilde \eta = \Xi_R(\eta \Lambda_L(m^{-1}(g,L)))^{-1},
\,\, m:= m_2 \circ m_1.
\label{K26}\ee
It is also not difficult to see that for any fixed $(g,b)$ the map
$\eta \mapsto \tilde \eta$ is a diffeomorphism of $\UN(n)$.
This leads to the following auxiliary statement.

\medskip
\noindent
{\bf Lemma 2.3.}
\emph{The actions $\tilde \cA^1$ and $\tilde \cA^2$ of $\UN(n)$ on
$\UN(n) \times \B(n)$ and on $\UN(n) \times \fP(n)$ defined by the formulas
\be
\tilde{\cA}_\eta^1(g,b) = (\eta g \eta^{-1}, \Lambda_L(\eta b)),
\qquad
\tilde{\cA}_\eta^2(g,L) = (\eta g \eta^{-1}, \eta L \eta^{-1}),
\quad \forall \eta\in \UN(n),
\label{tA2}\ee
have the same orbits as the respective Poisson actions $\cA^1$ and $\cA^2$.}

\medskip
It is plain from Lemma 2.3 that the tilded and the corresponding untilded
actions possess the same invariants.
On the other hand, for any Poisson action,
it is a standard fact that the Poisson bracket
of any two invariant functions is again invariant.
This leads to the next corollary.

\medskip
\noindent
{\bf Corollary 2.4.}
\emph{The ring of invariants $C^\infty(\UN(n) \times \fP(n))^{\UN(n)}$,
associated with the action $\tilde{\cA}^2$, is a Poisson subalgebra
of $C^\infty(\UN(n) \times \fP(n))$ with respect to the Poisson bracket
$\{\ ,\ \}_+^2$.}

\medskip
An analogous result holds for the model $\UN(n) \times \B(n)$
of the Heisenberg double as well.
We highlighted the statement of Corollary 2.4, since it will be used later.
Incidentally,  if a name is required at all, the action $\tilde \cA^2$ of $\UN(n)$ may be called
 \emph{undressed quasi-adjoint action}.

\section{Bi-Hamiltonian hierarchy on $T^*\UN(n)$}
\label{sec:2}

Let us consider the manifold
\be
\fM:= \UN(n) \times \Herm(n):= \{ (g, L)\mid g\in \UN(n), L\in \Herm(n)\},
\label{H1}\ee
which (as explained below) serves as a model of the cotangent bundle $T^* \UN(n)$.
Like in Section 2, for any function $F\in C^\infty(\fM)$,  we have the derivatives
\be
D_1F, D_1' F \in C^\infty(\fM, \b(n)) \quad\hbox{and}\quad d_2 F \in C^\infty(\fM, \u(n))
\label{H2}\ee
obeying the relation
\be
\langle  D_1 F(g,L), X\rangle +
\langle  D_1' F(g,L), X'\rangle + \langle d_2 F(g,L), Y\rangle
= \dt F(e^{tX}g e^{t X'}, L+ t Y),
\label{H3}\ee
 for every $X,X'\in \u(n)$ and $Y\in \Herm(n)$.

\medskip\noindent
{\bf Proposition 3.1.} \emph{The following formulas define two Poisson brackets on $C^\infty(\fM)$:
\be\{F, H\}_1(g,L) =
\left\langle D_1F , d_2H \right\rangle - \left\langle D_1H, d_2F \right\rangle
+ 2\left\langle L d_2F,  d_2H \right\rangle,
\label{H4}\ee
and
\bea
&&\{F, H\}_2(g,L) =  \left\langle D_1F , Ld_2H\right\rangle
-\left\langle D_1H, L d_2F \right\rangle
\nonumber \\
&&\qquad \qquad  + 2\left\langle L d_2F, \left(L d_2H \right)_{\u(n)}\right\rangle
-\frac{1}{2} \left\langle  D'_1F , g^{-1} (D_1H) g \right\rangle,
\label{H5}\eea
where the derivatives are taken at the point $(g,L)$ and we use the decomposition (\ref{Xdec}).}

\begin{proof}
The first bracket is the canonical Poisson bracket of the
cotangent bundle, expressed in terms of right-trivialization and taking $\Herm(n)=\ri \u(n)$ as the
model of $\u(n)^*$.  To see this,  note the identity
\be
2 \left\langle L d_2F, d_2 H\right\rangle = \left\langle L, [d_2 F, d_2 H]\right\rangle.
\label{H6}\ee
The restriction of the second bracket to the open submanifold $\UN(n) \times \fP(n) \subset \fM$
is a convenient multiple of the Heisenberg double Poisson  bracket (\ref{+PB2}).
Its algebraic nature guarantees that the Jacobi identity holds on the full manifold $\fM$.
For example, the Jacobi identity
\be
\{\{ L_a, L_b\}_2, L_c\}_2 + \hbox{c.p.}=0,
\ee
 for the linear functions $L_a(g,L):= \langle T_a, L\rangle$ defined
by a basis $\{T_a\}$ of $\u(n)$, requires the identity
\be
 \langle  L [T_a, (LT_b)_{\u(n)}] - L [T_b, (LT_a)_{\u(n)}] +[ LT_b, L T_a] ,
 (L T_c)_{\u(n)} \rangle + \hbox{c.p.} =0,
 \label{Jac}\ee
 where c.p. means cyclic permutations of the indices $a,b,c$.  Here, we used that
 \be
 d_2 L_a = T_a,\quad d_2 \{ L_a, L_b\}_2 = [T_a, (LT_b)_{\u(n)}] -  [T_b, (LT_a)_{\u(n)}] +
 (T_b L T_a - T_a L T_b),
\label{Hid}\ee
which is easily confirmed.
We know that the expression (\ref{Jac}) vanishes identically over the open subset
$\fP(n)\subset \fH(n)$,
because the Jacobi identity holds on the Heisenberg double.
Thus it vanishes  identically on the full $\fH(n)$, too, since it is given by a real
 analytic function of $L\in \fH(n)$.
The same argument holds for any three functions chosen from the $L_a$ and real and imaginary
parts of the matrix elements of $g$. This ensures the Jacobi identity
for all smooth functions, since the $L_a$ and some matrix elements of $g$ can always
be chosen locally as coordinate functions on $\fM$.
\end{proof}

Define the Hamiltonians
\be
H_k(g,L):= \frac{1}{k} \tr(L^k),\quad
\forall k \in \N.
\label{Hk}\ee
By using that $d_2 H_k = \ri L^{k-1}$, one deduces the next statement.

\medskip
\noindent
{\bf Proposition 3.2.}
\emph{The Hamiltonians $H_k$ pairwise Poisson commute with respect to both Poisson brackets
of Proposition 3.1, and satisfy the relation
\be
\{ F, H_k\}_2 = \{ F, H_{k+1}\}_1,
\qquad
\forall F\in C^\infty(\fM).
\label{biH}\ee
The flows of the  two Hamiltonian systems $\left(\fM,\{\ ,\ \}_2, H_k\right)$ and
$\left(\fM,\{\ ,\ \}_1, H_{k+1}\right)$
coincide, and are explicitly given by
\be
(g(t), L(t)) = \left( \exp( \ri t L(0)^k) g(0), L(0)\right).
\label{H9}\ee
}
\medskip

The flow of $(\fM, \{\ ,\ \}_1, H_1)$, given by $(g(t), L(t)) = (e^{\ri t} g(0), L(0))$, also commutes with
the above family.  We have a bi-Hamiltonian hierarchy, since the two Poisson brackets are
\emph{compatible},
i.e., their arbitrary linear combination is also a Poisson bracket. In order to show this,
thanks to well-known results (see e.g. \cite{FMP,Sm}), it is sufficient to prove Lemma 3.3 below.

Introduce the vector field $\cD$ on $\fM$ that acts as the following derivation of the evaluation
functions defined by
the matrix elements of $g$ and $L$:
\be
\cD[g_{ij}]:=0,
\quad
\cD[L_{ij}]:= \delta_{ij}.
\label{H10}\ee
Using the unit matrix  $\1_n$,
this is the vector field whose flow through $(g(0), L(0))$ reads
\be
(g(t), L(t)) = (g(0), L(0) + t \1_n).
\label{H11}\ee

\medskip
\noindent
{\bf Lemma 3.3.}
\emph{For $F\in C^\infty(\fM)$, let $\cD[F]$ denote the derivative along the vector field $\cD$.
The Poisson brackets of Proposition 3.1 enjoy the relation
\be
\{ F,H\}_1 = \cD[ \{ F,H\}_2] -\{ \cD[F], H\}_2 - \{ F, \cD[H]\}_2,
\label{D}\ee
which means that the first bracket is the Lie derivative of the second one.
In addition, we have
\be
\cD[ \{ F,H\}_1] -\{ \cD[F], H\}_1 - \{ F, \cD[H]\}_1 =0.
\label{D2}\ee
}

\begin{proof}
It is enough to check the relation (\ref{D}) for a set of coordinate
functions on $\fM$.  Let $L_a := \langle L, T_a\rangle$ be the component functions associated with a basis
$\{ T_a\}$ of $\u(n)$. The formula  (\ref{D}) certainly holds for coordinate functions on $\UN(n)$ and the $L_a$
if it holds for all elements of $C^\infty(\UN(n))$, which are regarded as $L$-independent elements
of $C^\infty(\fM)$, and all the functions $L_a$.
First, it is obvious that for $F,H\in C^\infty(\UN(n))$
both the left-hand side and the right-hand side of (\ref{D}) give zero.
Second, for $F\in C^\infty(\UN(n))$ and $H=L_a$ we get
\be
\cD[ \{ F,L_a\}_2] -\{ \cD[F], L_a\}_2 - \{ F, \cD[L_a]\}_2 = \cD[\{F, L_a\}_2] =
\langle D_1F, T_a\rangle = \{F, L_a\}_1.
\label{H13}\ee
Finally for $F=L_a$ and $H=L_b$, we obtain
\be
\cD[ \{ L_a,L_b\}_2] -\{ \cD[L_a], L_b\}_2 - \{ L_a, \cD[L_b]\}_2=
2\cD[\langle L T_a, (L T_b)_{\u(n)}\rangle]
= 2 \langle LT_a, T_b\rangle   = \{L_a, L_b\}_1,
\label{H14}\ee
and thus the proof of (\ref{D}) is complete. The equality (\ref{D2}) can be checked
along similar lines.
\end{proof}
\medskip

According to standard terminology \cite{FMP,Se,Sm}, $\fM=T^*\UN(n)$ equipped with the
two Poisson brackets subject to (\ref{D})and (\ref{D2})  is an example of
an exact bi-Hamiltonian manifold.
In conclusion, the identity  (\ref{biH}) shows that the Hamiltonians $H_k$ (\ref{Hk})
generate a bi-Hamiltonian
hierarchy on $\fM$.

\medskip
\noindent
{\bf Remark 3.4.}
The fact that the free geodesic motion on the Poisson-Lie group $\UN(n)$
corresponds to a Hamiltonian system on its Heisenberg double
was pointed out in \cite{Z}.
 Our bi-Hamiltonian description of the free hierarchy is apparently new.
 It is customary  to derive compatible
Poisson brackets by linearization of quadratic Poisson structures,
see, e.g., the paper \cite{RSk}.
Our construction is superficially similar, but we found
a compatible pair on the whole of $T^*\UN(n)$, whose existence is not implied
by general linearization arguments.

\section{Reduction under the conjugation action of $\UN(n)$}
 \label{sec:4}

The essence of  reduction with respect to a symmetry is that only
those observables of the physical system are kept that are invariant under the action
of the symmetry group. For the case at hand, this amounts to restriction to the ring of invariant
functions $C^\infty(\fM)^{\UN(n)}$, which
is customarily identified as $C^\infty(\fM/\UN(n))$.
Here, the invariance refers to the natural conjugation action of $\UN(n)$ on the cotangent bundle.
It can be viewed as an extension of the undressed quasi-adjoint action of Lemma 2.3, i.e.,
the action operates according to
\be
\tilde{\cA}_\eta^2(g,L) = (\eta g \eta^{-1}, \eta L \eta^{-1}),
\qquad
\forall \eta\in \UN(n),\, (g,L)\in \fM.
\ee
The reduction to invariant functions is often referred to as
\emph{Poisson reduction}. The following simple statement is important for us.

\medskip
\noindent
{\bf Lemma 4.1.} \emph{The Poisson brackets $\{\ ,\ \}_1$ and $\{\ ,\ \}_2$ of Proposition 3.1
induce two compatible Poisson brackets
on $C^\infty(\fM)^{\UN(n)}$.}
\medskip
\begin{proof}
This follows from the compatibility  of the two Poisson brackets on $\fM$ and from the fact
that the Poisson bracket of two smooth invariant functions is again invariant.
The latter fact is obvious for the first bracket and it is a known property (Corollary 2.4)
 of the restriction of the
second bracket to the Heisenberg double $\UN(n) \times \fP(n)$.
If $F$ and $H$ are $\UN(n)$-invariant real-analytic  functions, then the validity of the
invariance property,
\be
\{ F, H\}_2 \circ \tilde{\cA}_\eta^2 = \{ F, H\}_2,
\ee
over the open submanifold $\UN(n) \times \fP(n) \subset \fM$
implies that it holds over the full phase space $\fM$.
Indeed, both sides in the above equation represent real-analytic functions on $\fM$.
This ensures that the closure holds\footnote{Incidentally,
one can also work out a direct proof of the closure of $C^\infty(\fM)^{\UN(n)}$
under $\{\ ,\ \}_2$ (\ref{H5}).}
 for $C^\infty(\fM)^{\UN(n)}$, since \cite{Sch}  every smooth
invariant function on $\fM$ can be expressed as a smooth function of a finite set
of invariant polynomial functions
in the matrix elements of $g$ and $L$.
\end{proof}

We wish to study the  reduced Poisson algebras given by the Lemma 4.1.
In this paper, we make a technical assumption
that simplifies the required analysis.
Namely, we shall focus exclusively on the `regular part' of the phase space, and shall
characterize the Poisson brackets carried by $C^\infty(\fM_\reg)^{\UN(n)}$ (\ref{N2}).

Let $\T^n$ denote the standard maximal torus of $\UN(n)$. The dense open subset
$\T^n_\reg \subset \T^n$
contains the elements
 \be
 Q = \diag(Q_1,\dots, Q_n) \in \T^n \quad\hbox{for which}\quad Q_i \neq Q_j,\quad \forall i\neq j.
\label{N1} \ee
The dense open subset $\UN(n)_\reg \subset \UN(n)$ is filled by the conjugacy classes
passing through $\T^n_\reg$.
We define
\be
\fM_\reg:= \UN(n)_\reg \times \Herm(n).
\label{N2}\ee
Every $\UN(n)$ orbit in $\fM_\reg$ contains representatives in the submanifold
\be
\T^n_\reg \times \Herm(n) \subset \fM_\reg,
\label{N3}\ee
and this submanifold is preserved by the action of the  normalizer, denoted $\cN(n)$,
of $\T^n$ in $\UN(n)$,
\be
\cN(n) \equiv \{ \eta \in \UN(n)\mid \eta Q \eta^{-1} \in \T^n,\quad \forall Q\in \T^n\}.
\label{N4}\ee
The ring of the $\cN(n)$-invariant functions on  $\T^n_\reg \times \fH(n)$
will serve as a model of $C^\infty(\fM_\reg)^{\UN(n)}$.

\medskip
\noindent
{\bf Lemma 4.2.}
\emph{Let $\iota: \T^n_\reg \times \Herm(n) \to \UN(n)_\reg \times \Herm(n)$ be the
tautological embedding.
For any $F\in C^\infty(\fM_\reg)^{\UN(n)}$,  $\iota^* F \equiv F \circ \iota$ belongs to
$C^\infty(\T^n_\reg \times \Herm(n))^{\cN(n)}$, and
\be
\iota^*: C^\infty(\fM_\reg)^{\UN(n)} \to C^\infty(\T^n_\reg\times \Herm(n))^{\cN(n)}
\label{N5}\ee
is an isomorphism of commutative algebras.}

\begin{proof}
Any $F\in C^\infty(\fM_\reg)^{\UN(n)}$ is uniquely determined by its restriction to
$\T^n_\reg \times \Herm(n)$,
and the restricted function belongs to $C^\infty(\T^n_\reg \times \Herm(n))^{\cN(n)}$.
For any $g\in \UN(n)_\reg$, we can choose $\sigma_1(g)\in \T^n_\reg$ and
$\sigma_2(g) \in \UN(n)$ such that
$\sigma_1(g) = \sigma_2(g) g \sigma_2(g)^{-1}$.
Then, for any $f\in  C^\infty(\T^n_\reg\times \Herm(n))^{\cN(n)}$, the formula
\be
F(g,L) := f (\sigma_1(g), \sigma_2(g) L \sigma_2(g)^{-1})
\label{Fdef}\ee
gives a well-defined, $\UN(n)$-invariant function on $\fM_\reg$.
To see that this function is smooth, we note that $\UN(n)_\reg$ is the base
of the (left) $\cN(n)$ principal fibre bundle
with total space
$\T^n_\reg \times \UN(n)$, $\cN(n)$ action given by
$\cN(n)\ni \nu: (\tau, \eta) \mapsto (\nu \tau \nu^{-1}, \nu \eta)$,
and bundle projection
 $\T^n_\reg \times\UN(n) \ni (\tau, \eta) \mapsto  \eta^{-1} \tau \eta \in \UN(n)_\reg$.
Since this bundle is locally trivial\footnote{
We here use some well-known results about free proper actions; see, e.g., paragraph 6.5 in \cite{Rud}.},
it admits smooth local sections,
\be
g\mapsto \sigma(g) = (\sigma_1(g), \sigma_2(g)).
\ee
Using such section $\sigma$ in (\ref{Fdef}) shows that $F$ is locally smooth.
Because $F$ is a globally well-defined function on $\fM_\reg$,
we see that it  belongs to $C^\infty(\fM_\reg)$.
\end{proof}

\medskip
\noindent
{\bf Definition 4.3.}
The reduced Poisson algebras
$\left(C^\infty(\T^n_\reg \times \Herm(n))^{\cN(n)}, \{\ ,\ \}_i^\red\right)$ are defined by setting
\be
\{ F \circ \iota , H\circ \iota \}_i^\red := \{ F, H\}_i \circ \iota\quad
\quad\hbox{for}\quad
F, H \in C^\infty(\fM_\reg)^{\UN(n)},\,\, i=1,2.
\label{N6}\ee
\medskip

We shall establish  an intrinsic description  of the reduced Poisson brackets (\ref{N6}).
In preparation,
let us decompose $\gl$ as the direct sum of subalgebras
 \be
 \gl= \gl_+ + \gl_0 + \gl_-
 \label{N7}\ee
 by means of the principal gradation, i.e., $\gl_0$ contains the diagonal matrices, and
  $\gl_+$ ($\gl_-$)  contains
the  strictly upper (lower) triangular matrices.
 Correspondingly,  any $X\in \gl$ can be written in the form $X=X_+ + X_0 + X_-$.
We may also write $X= X_0 + X_\perp$ with $X_\perp := X_+ + X_-$.

 For $Q\in \T_\reg^n$, the linear operators
 $(\Ad_Q - \id)\vert_{\gl_{\pm}}$ are invertible,
 and therefore one may introduce $\cR(Q)\in \mathrm{End}(\gl)$ by setting it equal to zero on $\gl_0$ and
 defining it otherwise as
 \be
 {\cR(Q)\vert}_{\gl_+ + \gl_-} =
 \frac{1}{2}(\Ad_Q + \id) \circ \left({(\Ad_Q - \id)\vert}_{\gl_+ + \gl_-}\right)^{-1},
 \label{N8}\ee
 where $\Ad_Q(X)= Q X Q^{-1}$ for all $X\in \gl$.
 Incidentally, this is a well-known solution of the modified classical dynamical
 Yang-Baxter equation \cite{EV}, which first appeared in \cite{BDF}.
 Below, we apply the notation
 \be
 [X,Y]_{\cR(Q)} := [ \cR(Q) X, Y] + [X, \cR (Q)Y],
 \quad\forall X,Y\in \gl.
 \label{N9}\ee

 For  any $f\in C^\infty(\T^n_\reg \times \Herm(n))$,  the
$\b(n)_0 := \b(n)\cap \gl_0$-valued derivative $D_1 f$ and the $\u(n)$-valued derivative $d_2 f$
are defined naturally,  in analogy with  (\ref{H3}):
\be
\langle  D_1 f(Q,L), X\rangle
+ \langle d_2 F(Q,L), Y\rangle
= \dt f(e^{tX}Q , L+ t Y),
\label{Dfdef}\ee
 for every $X\in \u(n)_0:= \u(n)\cap \gl_0$ and $Y\in \Herm(n)$.
 Here, $D_1f$ is well-defined because $e^{tX} Q \in \T^n_{\reg}$ for $t$ small enough.

\medskip
\noindent
{\bf Lemma 4.4.}
\emph{Let $f := F \circ \iota$ for a function $F\in C^\infty(\fM_\reg)^{\UN(n)}$.
Then the following relations hold at any $(Q,L) \in \T^n_\reg \times \Herm(n)$:
\be
d_2 F(Q,L) = d_2 f(Q,L),\quad
[L, d_2 f(Q,L)]_0 = 0,
\label{N11}\ee
\be
D_1 F(Q,L) = D_1 f(Q,L) - [L, d_2 f(Q,L)]_+  - 2 \cR(Q)[ L, d_2 f(Q,L)]_+,
\label{N12}\ee
where the subscripts $0$ and $+$ refer to the decomposition (\ref{N7}).}
\medskip
\begin{proof}
The first equality in (\ref{N11}) is trivial, and the second one follows from
\bea
&&0=\dt f(Q, e^{tX} L e^{-tX}) =\dt f(Q, L + t [X,L] + o(t)) \nonumber \\
&&\quad = \langle d_2 f(Q,L), [X,L]\rangle
= \langle [L, d_2 f(Q,L)]_0, X\rangle,
\quad
\forall X \in \u(n)_0.
\eea
In order to derive (\ref{N12}), let us take an arbitrary off-diagonal element $T\in \u(n)$,
and use the invariance of $F$ together with the first equality in (\ref{N11})
 to write
\be
0 = \dt F(e^{tT} Q e^{-tT}, e^{tT} L e^{-tT}) = \langle T, D_1 F(Q,L) - D_1' F(Q,L) +
[L, d_2 f(Q,L)] \rangle.
\label{rel18}\ee
On account of the relation $D_1'F (Q,L) = \Ad_Q^{-1} ( D_1F(Q,L))$ and some obvious identities,
equation (\ref{rel18}) is   equivalent to
\be
0=\langle T_-, (\Ad_Q - \id) \circ \Ad_Q^{-1} (D_1 F(Q,L))_+ + 2 [L,d_2 f(Q,L)]_+\rangle.
\ee
To get this, we noticed that, with the decomposition  $T=T_- + T_+$, we have
\be
\langle T, [L, d_2 f(Q,L)] \rangle = 2 \langle T_-, [L, d_2 f(Q,L)]\rangle.
\ee
As a result, we see that (\ref{rel18}) is   equivalent to
\be
(\Ad_Q - \id) \circ \Ad_Q^{-1} (D_1 F(Q,L))_+ = - 2 [L,d_2 f(Q,L)]_+.
\ee
Since $Q$ is regular, this can be solved for $(D_1 F(Q,L))_+$.
Using that the inverse is well-defined on $\gl_+$,
the solution is
\be
(D_1 F(Q,L))_+ = - 2 \Ad_Q \circ (\Ad_Q - \id)^{-1} [L, d_2 f(Q,L)]_+ =
- (\id +  2 \cR(Q)) [L,d_2f(Q,L)]_+.
\ee
Combining this with the equality $(D_1 F(Q,L))_0 = D_1 f(Q,L)$,
which is a direct consequence of the definitions,
one obtains the claimed formula (\ref{N12}).
\end{proof}
\medskip

The main result of this  paper is
the following description of
the reduced Poisson brackets.

\medskip
\noindent
{\bf Theorem 4.5}.
\emph{For $f, h\in C^\infty(\T_\reg^n \times \Herm(n))^{\cN(n)}$, the reduced Poisson brackets
(\ref{N6}) obey the explicit formulas
\be
\{ f, h\}_1^\red(Q,L) =
\langle D_1 f, d_2 h\rangle - \langle D_1 h, d_2 f\rangle +\langle L,
[ d_2 f, d_2 h]_{\cR(Q)}\rangle,
\label{N13}\ee
and
\be
\{ f, h\}_2^\red(Q,L) =
\langle D_1 f, L d_2 h\rangle - \langle D_1 h, L d_2 f\rangle +
2\langle L d_2 f, \cR(Q) (L d_2 h)\rangle.
\label{N14}\ee
The derivatives are evaluated at the point $(Q,L)$, and the notations (\ref{N8}), (\ref{N9})
 are applied.}
\medskip
\begin{proof}
As detailed below, the claimed formulas result by  substituting the formulas of Lemma 4.4 into the
Poisson bracket formulas of  Proposition 3.1, and performing some elementary algebraic manipulations.

To deal with the first Poisson bracket, note that at the point $(Q,L)$ we have
\be
- 2 \langle \cR(Q) [L, d_2 f]_+, d_2 h\rangle = 2 \langle [L, d_2 f]_+, \cR(Q) (d_2 h)\rangle =
\langle L, [d_2 f, \cR(Q) (d_2h)]\rangle.
\ee
To get this, we used the anti-symmetric nature of $\cR(Q)$  together with the fact that
it maps $\u(n)$ to $\u(n) \cap \left(\gl_+ + \gl_-\right)$, and the obvious
identity $\langle X, Y\rangle =
- \langle X^\dagger, Y^\dagger\rangle$.
We also have
\be
- \langle [L, d_2 f]_+, d_2 h] \rangle = - \frac{1}{2} \langle [L, d_2 f]_\perp, d_2 h\rangle =
 -\frac{1}{2} \langle L, [d_2 f, d_2 h]\rangle,
\ee
where the last equality crucially depends on the property $[L, d_2 f]_0 =0$.
The formula (\ref{N13}) results by using these relations and Lemma 4.4 for the evaluation of
\be
\{ f, h\}_1^\red = \langle (D_1 F)_0 + (D_1 F)_+, d_2 H \rangle - \langle (D_1 H)_0 +
(D_1 H)_+, d_2 F \rangle
+ \langle L, [d_2 F, d_2 H]\rangle
\ee
at the point $(Q,L)$.

Turning to the derivation of (\ref{N14}), at the point $(Q,L)$, we record the identities
\bea
&&\langle D_1 F, L d_2H \rangle = \langle D_1 f, L d_2 h \rangle -
\langle (L d_2 h)_{\u(n)}, \frac{1}{2} [L, d_2 f] + \cR(Q) [L, d_2f]\rangle
\nonumber \\
&& \qquad = \langle D_1 f, L d_2 h \rangle +  \langle 2 \cR(Q)(L d_2h)_{\u(n)} -
(L d_2h)_{\u(n)}, L d_2 f\rangle,
\eea
\be
2 \langle L d_2 F, (L d_2 H)_{\u(n)} \rangle = \langle L d_2 f, (L d_2 h)_{\u(n)} \rangle -
 \langle L d_2 h, (L d_2 f)_{\u(n)} \rangle.
\label{N29}\ee
\be
\langle D_1' F, Q^{-1} (D_1 H) Q \rangle =0.
\ee
To verify (\ref{N29}), notice that $\Im\tr(L (d_2F) L (d_2H)) = 0$, because
$\Im \tr(X^\dagger) = - \Im \tr(X)$  for all $X\in \gl$,
then use the decomposition (\ref{Xdec}).
With the aid of these identities, equation (\ref{H5}) gives
\bea
&&\{ f, h\}_2^\red(Q,L) =
\langle D_1 f, L d_2 h\rangle - \langle D_1 h, L d_2 f\rangle \nonumber\\
&&\qquad
+ 2\langle L d_2 f, \cR(Q) (L d_2 h)_{\u(n)}\rangle
-  2\langle \cR(Q)(L d_2 f)_{\u(n)},  L d_2 h\rangle.
\label{preN14}\eea
Finally, noting the identity
\be
-  \langle \cR(Q)(L d_2 f)_{\u(n)},  L d_2 h\rangle =
\langle (L d_2 f)_{\u(n)}, \cR(Q) (L d_2 h)\rangle =
 \langle L d_2 f , \cR(Q) (L d_2 h)_{\b(n)}\rangle,
\ee
equation (\ref{preN14}) is converted into (\ref{N14}).
\end{proof}

Now we describe the reduction of the equations of motion of the bi-Hamiltonian
hierarchy (\ref{biH}).
Denote by $V_k$ the bi-Hamiltonian vector field on $\fM$ satisfying
\be
V_k[F] = \{ F, H_k\}_2 = \{ F, H_{k+1}\}_1, \qquad k\in \N.
\label{N15}\ee
This induces a derivation of $C^\infty(\fM)^{\UN(n)}$, which in turn translates into
a derivation of $C^\infty(\T^n_\reg \times \Herm(n))^{\cN(n)}$.
The latter derivation corresponds to a (non-unique) vector field $W_k$ on the manifold
$\T^n_\reg \times \Herm(n)$, whose value at $(Q,L)$ takes the following form:
\be
W_k(Q,L) = V_k(Q,L) + \left([\zeta(Q,L), Q], [\zeta(Q,L), L]\right),
\label{N16}\ee
where $V_k(Q,L) = (\ri L^k Q, 0)$, according to (\ref{H9}),   and $\zeta(Q,L)\in \u(n)$
 is subject to the condition
\be
(\ri L^k Q + [\zeta(Q,L),Q])Q^{-1} \in \u(n)_0.
\label{N17}\ee
In words, the `infinitesimal gauge transformation' $\zeta(Q,L)$ ensures that $W_k$
is tangential to the manifold $\T^n_\reg \times \Herm(n)$.
This holds since $\u(n)_0 = \u(n) \cap \gl_0$ is the Lie algebra of $\T^n$.

\medskip
\noindent
{\bf Proposition 4.6.}
\emph{The induced evolutional vector field $W_k$ of Eq.~(\ref{N16}) is given by
\be
W_k(Q,L) = \left (i (L^k)_0 Q, [\cR(Q)(\ri L^k), L] \right),
\label{N18}\ee
up to an arbitrary function $\delta \zeta_0(Q,L) \in \u(n)_0$ that does not effect
the induced derivatives
of the elements of $C^\infty(\T^n_\reg\times \Herm(n))^{\cN(n)}$.
By choosing $\delta \zeta_0 =0$,  the evolution equation for
$(Q(t), L(t)) \in \T^n_\reg \times \fH(n)$ associated  with
the vector field $W_k$ has the form (\ref{I1}).
}
\medskip
\begin{proof}
One can check that
\be
\zeta(Q,L) = \cR(Q)(\ri L^k) - \frac{\ri}{2} L^k
\label{N19}\ee
is a particular solution of the condition (\ref{N17}).
The general solution is obtained by adding $\delta\zeta_0$ to this one.
Substitution of (\ref{N19}) into (\ref{N16}) gives (\ref{N18}).
\end{proof}

\medskip

To sum up,
the message of
Proposition 4.6 is that \emph{our Poisson reduction yields the
 trigonometric spin Ruijsenaars--Sutherland hierarchy} as defined in Section 1.
The reduction treatment  equips this hierarchy with a bi-Hamiltonian structure.
Indeed, our construction implies that the evolutional derivatives of the gauge
invariant observables
 $f \in C^\infty(\T^n_\reg \times \fH(n))^{\cN(n)}$ satisfy
 \be
 W_k[f] = \{ f, h_k\}_2^\red = \{ f, h_{k+1}\}_1^\red,
 \ee
 with the compatible Poisson brackets given by Theorem 4.5
 and the reduced Hamiltonians $h_k$ obtained from $H_k$ (\ref{Hk}).
 We next present a large set of
constants of motion for  this hierarchy.

\medskip
\noindent
{\bf Proposition 4.7.}
\emph{Let $\cP( L, Q^{-1} L Q)$ be an arbitrary `non-commutative polynomial' , i.e.,
a linear combination of ordered products
of powers of $L$ and $Q^{-1}  L Q$. Then the $\cN(n)$-invariant function
$\tr \left(\cP(L, Q^{-1}  L Q)\right)$
is constant along the flow of the evolutional vector field $W_k$.}

\medskip
\noindent
\begin{proof}
Denoting the derivative along $W_k$ by $\frac{d}{dt}$, we observe that
\be
\frac{d}{dt} Q = \ri L^k  Q + [\zeta(Q,L), Q]
\quad\hbox{and}\quad
\frac{d}{dt}L= [\zeta(Q,L), L]
\label{N21}\ee
imply
\be
\frac{d}{dt}(Q^{-1} L Q) = [\zeta(Q,L), Q^{-1}LQ].
\label{N22}\ee
This in turn implies that the evolutional derivative of $\cP(L, Q^{-1} L Q)$
has the form
\be
\frac{d}{dt}\cP( L, Q^{-1} L Q)= [\zeta(Q,L), \cP( L, Q^{-1}L Q)],
\label{N23}\ee
and thus the derivative of $\tr(\cP)$ is zero.
The $\cN(n)$-invariance of these conserved quantities
follows from the invariance of the trace with respect to conjugations.
\end{proof}

Proposition 4.7 provides constants of motion for the reduced bi-Hamiltonian dynamics
on $\fM^\reg/\UN(n)$.
These constants of motion  are restrictions of well-defined functions on the whole
of $\fM/\UN(n)$.
Indeed, the formula
\be
\tr\left( \cP(L, g^{-1} L g)\right)
\label{N24}\ee
gives a  $\UN(n)$-invariant, smooth function of $(g,L)\in \fM$,
which is a constant of motion for all the bi-Hamiltonian vector fields displayed in
 equation (\ref{biH}).
It reproduces  $\tr \left(\cP(L, Q^{-1}  L Q)\right)$ upon restriction to
$\T^n_\reg \times \fH(n)$.
 The Poisson brackets and the algebraic relations of these constants of motion will be
 studied in a future publication.

The proper analogues of Proposition 4.6 and Proposition 4.7 hold
for the reduction of a suitable free system on the Heisenberg double of any compact simple
 Lie group
as well \cite{F1}, but (at least at present) we do not have a bi-Hamiltonian structure in
such general case.

We end by remarking that arguments similar to those utilized  by Reshetikhin \cite{Res1,Res2}
can be applied to show the degenerate integrability of the reduced dynamics
on generic symplectic leaves of any of the two reduced Poisson brackets.
However, the details are rather complicated since $\fM/ \UN(n)$ is not a smooth manifold.
This issue should be investigated further invoking the machinery of singular Hamiltonian
reduction \cite{OR,SL}.
One of the interesting open questions is whether the above exhibited polynomial constants
of motion are sufficient for the degenerate integrability of the reduced system.

\section{Discussion}

The first new result of this paper is the bi-Hamiltonian description of the free motion on the group $\UN(n)$,
developed in Section 3.
We noticed that it is useful to present the Poisson structure \cite{STS} of the
Heisenberg double
in terms of the variables $(g_R, b_R b_R^\dagger)\in \UN(n)\times \fP(n)$,
since in this way it admits extension to the cotangent bundle $\fM = \UN(n) \times \fH(n)$.
In Section 4, we  demonstrated that Poisson reduction
of the hierarchy of free motion leads to
the trigonometric spin Ruijsenaars--Sutherland  hierarchy governed by the evolution
 equations (\ref{I1}).
This yields a bi-Hamiltonian interpretation for the dynamics of the gauge invariant functions
of the variables $(Q,L)$,  where the gauge group is given by the normalizer $\cN(n)$ of the maximal
torus $\T^n$ inside $\UN(n)$.

The interpretation of the reduced system as a spin Sutherland model is supported by the
the change of variables (\ref{I5}) that brings the Hamiltonian
$\frac{1}{2} \tr (L^2)$ into the form (\ref{I7}), and  converts the reduced
first Poisson bracket into the natural one carried by the phase space
$\left(T^* \T^n_\reg \times  (\u(n)^*//_0\T^n)\right)/ S_n$.

We now briefly discuss another change of variables, which is suited for the second
 reduced Poisson bracket upon restriction
to the open submanifold arising from the Heisenberg double
$\UN(n)\times \fP(n) \subset \fM$.
In this case we can  write $L=bb^\dagger$, where
$b= e^p b_+$ with a real diagonal matrix $p\in \b(n)_0$ and an upper triangular matrix
having unit diagonal, $b_+ \in \B(n)_+$.
By introducing $\lambda:= b_+^{-1} Q^{-1}   b_+ Q$, we obtain the
invertible change of variables
\be
\T^n_\reg \times \fP(n)\ni (Q, L) \longleftrightarrow
(Q,p, \lambda) \in \T^n_\reg \times \b(n)_0 \times \B(n)_+,
\label{F1}\ee
whereby every function $f(Q,L)$ is represented by a function $\cF(Q,p,\lambda)$.
It can be shown (both by direct calculation or by applying Theorem 4.3 of \cite{F1})
that the reduced
second Poisson bracket acquires the following decoupled form in terms of the new variables:
\be
2 \{\cF, \cH\}^\red_2(Q,p,\lambda) = \langle D_Q \cF, d_p \cH \rangle -
\langle D_Q \cH, d_p \cF \rangle
+ \langle  D'_\lambda \cF, \lambda^{-1} (D_\lambda \cH) \lambda \rangle.
\label{F2}\ee
The derivatives on the right hand side are taken at $(Q,p,\lambda)$,
$D_Q\cF \in \b(n)_0$ and $d_p \cF = \u(n)_0$ are defined in the obvious manner, and
we  take $D_\lambda \cF $ and $D'_\lambda \cF $ from the off-diagonal subspace of $\u(n)$,
 according to the rule
 \be
\langle  D_\lambda \cF(Q,p,\lambda), X_+\rangle  + \langle D'_\lambda \cF(Q,p,\lambda),
Y_+\rangle
= \dt \cF(Q,p, e^{tX_+}\lambda  e^{t Y_+}),
\label{F3}\ee
 $\forall  X_+, Y_+\in \b(n)_+$.  The subgroup $\T^n$ of the gauge group acts
 by $(Q,p,\lambda)\mapsto  (Q,p, \tau \lambda \tau^{-1})$,
 and the formula (\ref{F2}) defines a Poisson bracket on the $\T^n$-invariant functions.
 Its last term can be recognized as the natural reduced Poisson
 bracket on $\B(n)//_0 \T^n$, which is the Poisson-Lie analogue of $\u(n)^*//_0\T^n$.
 The Hamiltonian $\tr(L)$ has the `spin Ruijsenaars form'
 \be
 \tr(L) = \sum_{i=1}^n e^{2p_i} V_i(Q,\lambda) \quad\hbox{with}\quad
 V_i(Q,\lambda)=\left(b_+(Q,\lambda) b_+(Q,\lambda)^\dagger\right)_{ii},
 \label{F4}\ee
where $\lambda$ represents a `spin' variable.
An enlightening explicit formula  of  $V_i(Q,\lambda)$ is not available in
general\footnote{A complicated
explicit formula for $b_+(Q,\lambda)$ can be obtained along the lines of
Section 5.2 in \cite{F1}.}, but
it is known that restriction to a particular symplectic leaf
 of $\B(n)//_0 \T^n$ gives the spinless trigonometric RS model \cite{FK1}.
An unpleasant feature of the new variables $(Q,p,\lambda)$ is that the action
of the full gauge group $\cN(n)$, and that of the permutation group $S_n = \cN(n)/\T^n$,
 is not transparent in this setting
 (for the spinless case, see Section 4 of \cite{FK2}).

There is a link between our results and
the observation of Suris \cite{Sur}, who noticed
that the \emph{spinless} RS and Calogero--Moser hierarchies are governed by the
same $R$-operators. In the trigonometric case, the pertinent $R$-operator is the sum of the one
in (\ref{I2}) and a `correction term'.  In this case the statement of \cite{Sur} can be
derived from our results by applying suitable restrictions and gauge fixings to the
spin Ruijsenaars--Sutherland hierarchy.

An interesting open problem that stems from our work is that
 the global structure of the full reduced phase space should be explored in the future,
 dropping the restriction to $\fM_\reg \subset \fM$.
 The issue of possible
 generalizations of the bi-Hamiltonian structure to the elliptic case and for other Lie groups should be also investigated.
 Finally, we wish to mention the question whether there is any relation between our results
 and the earlier studies \cite{Bar,FaMe}
of a bi-Hamiltonian structure for the rational Calogero--Moser system.

\bigskip
\bigskip
\begin{acknowledgements}
 I wish to thank I. Marshall for generous help with some calculations.
 I am also grateful to J. Balog, T.F. G\"orbe and M. Fairon for useful comments on the manuscript.
 This research was performed  in the framework of the project
GINOP-2.3.2-15-2016-00036 co-financed by the European Regional
Development Fund and the budget of Hungary.
\end{acknowledgements}


\end{document}